\documentclass[aps,pra,preprint,showpacs]{revtex4}
\usepackage{bm,graphicx,amsmath}
\begin{document}
\title{State-dependent rotations of spins by weak measurements}
\author{D. J. Miller}
\email[]{davmille@arts.usyd.edu.au}
\affiliation{Centre for Time, University of Sydney, Sydney NSW 2006, Australia and School of Physics, University of New South Wales, Sydney NSW 2052, Australia}
\date{\today}

\begin{abstract}
It is shown that a weak measurement of a quantum system produces a new state of the quantum system which depends on the prior state, as well as the (uncontrollable) measured position of the pointer variable of the weak measurement apparatus. The result imposes a constraint on hidden-variable theories which assign a different state to a quantum system than standard quantum mechanics. The constraint means that a crypto-nonlocal hidden-variable theory can be ruled out in a more direct way than previously.
\end{abstract}

\maketitle

\section{Introduction \label{intro}}

In a standard von Neumann measurement, or normal measurement (NM), the state of the quantum system is transformed to one of the eigenstates of the observable $A$ that is being measured. In a weak measurement (WM) \cite{av1,av2} of $A$, the state of the quantum system is transformed to a linear combination of the eigenstates of $A$.  For a measurement of spin, which we will be concerned with here, the transformation of state as a result of either type of measurement can be visualized as a rotation of the direction of spin. In Sect.~II, it is shown that the new spin-direction as a result of a WM depends on both the spin direction of the quantum system before measurement (the ``prior" state) and the pointer position which records the result of the WM.

Since the new spin-direction depends on the spin direction of the prior state, it is interesting to apply the result to WMs of maximally entangled spin-states. This is because in standard quantum mechanics the prior state is the reduced state of each of the entangled subsystems which does not specify a direction in space for the subsystem's spin. On the other hand, some hidden-variable (HV) theories of quantum mechanics do assign spin states to the subsystems of an entangled pair so the spin-direction dependent WM rotation provides an additional means of testing such theories. Leggett  \cite{leggett} has shown that a nonlocal HV theory of a certain type (a ``crypto-nonlocal" (CNL)  theory) is ruled out by a different set of inequalities from the Bell inequalities \cite{bell}. The new inequalities have been tested experimentally \cite{groblacher,branciard}. In Sec.~IIIB, the rotations due to WMs are used to give a more direct demonstration that the CNL HV theory of Ref.~\onlinecite{leggett} is not viable.

\section{Weak measurements}

The concept of a WM was introduced by Aharonov and co-workers (for a discussion, see Section 3 of Ref.~[\onlinecite{av2}]). A WM involves a ``weak" (in the sense described below) interaction between a measuring instrument, in this case the weak measurement apparatus (WMA), and the quantum system.  The WM outcome is recorded by a NM of the ``pointer'' position $Q$ of the WMA. 

The observable of the quantum system which is to be measured is the Pauli spin operator $\hat{\sigma}_z$, twice the component of the spin of the quantum system along the $z$-axis (which is a direction set by the WMA) in units of $\hbar$. The interaction Hamiltonian for the WM  is
\begin{eqnarray} \label{H}
\hat{H}(t) & = & a g(t)  \hat{\sigma_z} \hat{P}.
\end{eqnarray}
Here $\hat{P}$ is the operator which is conjugate to $\hat{Q}$ and the constant $a$ ($-a$) is the average distance the WM pointer moves when a quantum system with spin up (down) along the $z$-axis interacts with the WMA. We assume that, except for the interaction, the quantum system and the WMA evolve freely, meaning that $\hat{\sigma}_z$ and $\hat{P}$ are constants of the motion between the preparation at $t_0$ of the quantum system and the WMA and the time when the NM of the WMA pointer-position is made at $t_{WM}$ (the interaction between the quantum system and the WMA may cease prior to $t_{WM}$). The real, scalar function of time $g(t)$ is normalized, $\int_{t_{0} }^{t_{WM}} g(t) dt = 1$.  The only relevant factor of the time-evolution operator is
\begin{equation} 
\label{U}
\hat{U}(t_{WM},t_{0}) = \exp  \left[- \frac{i }{\hbar} {\int_{t_{0}}^{t_{WM}} a g(t)  \hat{\sigma_z} \hat{P} dt}\right]= \exp\left[{-i \frac{a}{\hbar} \hat{\sigma_z}\hat{P}}\right]. 
\end{equation} 

The WMA is prepared in the state $|\psi;t_0 \rangle$ at time $t_0$ with $\langle \psi;t | \hat{P} |\psi;t \rangle = 0$ for $t_0<t<t_{WM}$, so that the pointer does not drift with time, and the origin in space can be chosen so that $\langle \psi;t_0 | \hat{Q}|\psi;t_0\rangle = 0$. For a WM, one requires that $a \lesssim \Delta_{\psi} = \sqrt{\langle \psi | \hat{Q}^2|\psi\rangle-\langle \psi | \hat{Q}|\psi\rangle^2}$, the width of the initial distribution of pointer positions. Thus when the WM is completed by observation (or decoherence) of the pointer position, the quantum system is not unambiguously left in one of the eigenstates of  $\hat{\sigma_z}$ as it is for a NM for which $a \gg \Delta_{\psi}$. 

The effect of the time evolution on the spin states $|\pm \rangle$ of the quantum system along the $z$-axis and the state $|\psi\rangle$ of the WMA is
\begin{eqnarray} 
e^{-i \frac{a}{\hbar} \hat{\sigma}_z\hat{P}} |\pm \rangle |\psi \rangle
& = &|\pm \rangle \hat{S}(\pm a) |\psi \rangle \label{gaussian}
\end{eqnarray}
where $\hat{S}(\lambda)$ is the translation operator \cite{qmbook} in the Hilbert space of the WMA pointer with the property $\hat{S}(\lambda)|Q\rangle = |Q+\lambda \rangle$.

\subsection{Case of a single spin}

Consider a single spin prepared in the state 
\begin{equation}
\label{eta}
|p_+ \rangle = \cos \frac{\theta_{p}}{2} e^{-i\phi_{p}/2} |+\rangle + \sin \frac{\theta_{p}}{2} e^{i\phi_{p}/2} |- \rangle,
\end{equation} 
i.e. with spin up in a direction $\hat{\bm{p}}$ specified by the polar coordinates $\theta_{p},\phi_{p}$, making an angle $\theta_{p}$ with the $z$-axis (specified by the WMA) and $\phi_{p}$ with the $x$-axis (in this case chosen arbitrarily). After the interaction and an observation identifying the WMA pointer-position to be $Q=Q_1$, the normalized state of the quantum system and WMA is 
\begin{eqnarray} 
\label{alphaspin}
&&\frac{1}{\sqrt{N}}| Q_1\rangle \langle Q_1|e^{-i \frac{a}{\hbar} \hat{\sigma}_z\hat{P}}|p_+ \rangle |\psi;t_{0} \rangle \nonumber \\
& = & \frac{1}{\sqrt{N}}\left[\langle Q_1| \hat{S}(a)|\psi;t_{0}\rangle \cos \frac{\theta_{p}}{2} e^{-i\phi_{p}/2} |+ \rangle + \langle Q_1| \hat{S}(-a)|\psi;t_{0} \rangle \sin \frac{\theta_{p}}{2} e^{i\phi_{p}/2} |- \rangle \right] | Q_1\rangle  \nonumber \\
& = & \frac{1}{\sqrt{N}}\left[\psi(Q_1-a,t_{WM}) \cos \frac{\theta_{p}}{2} e^{-i\phi_{p}/2} |+ \rangle + \psi(Q_1+a,t_{WM}) \sin \frac{\theta_{p}}{2} e^{i\phi_{p}/2} |- \rangle \right] | Q_1\rangle 
\end{eqnarray}
where $\psi(Q,t) = \langle Q| \psi;t \rangle$ is the wave function of the WMA pointer and the normalization constant $N = |\psi(Q_1-a,t_{WM})|^2 \cos^2 \frac{\theta_{p}}{2} + |\psi(Q_1+a,t_{WM})|^2 \sin^2 \frac{\theta_{p}}{2}$.  

Thus, as a result of the WM, the spin of the quantum system is up in a direction $\hat{\bm{q}}$ specified by the polar angles $\theta_q$ and $\phi_q$ where
\begin{eqnarray}
\tan \frac{\theta_{q}}{2} & = & |f(Q_1)| \tan \frac{\theta_{p}}{2} \label{eq;tangamma} \\
\phi_{q} & = & \phi_{p} + \arg f(Q_1)
\end{eqnarray}
where
\begin{equation} 
f(Q_1)=\frac{\psi(Q_1+a,t_{WM})}{\psi(Q_1-a,t_{WM})}.
\end{equation}
Throughout the following we will assume that $\phi_{q}=\phi_{p}$ and independent of time, which is a good approximation if, for example, the initial distribution of $Q$-values is Gaussian and is sufficiently broad so that further broadening with time between $t_0$ and $t_{WM}$ can be neglected. Under those conditions $f(Q_1)$ is real and positive.

As a result of the WM, the spin of the quantum system is rotated through an angle $\Delta \theta (\theta_{p}, Q_1)$ away from the $z$-axis where
\begin{equation}
\label{eq;dthetaf}
\Delta \theta (\theta_{p}, Q_1) =  2 \arctan \left[f(Q_1) \tan \frac{\theta_{p}}{2}\right] - \theta_{p} 
\end{equation}
which depends on both the initial angle $\theta_{p}$ between the spin-direction and the direction of the positive $z$-axis set by the WMA and the (uncontrollable) result $Q_1$ of the WM. Note that if $f(Q_1)<1$, the spin is rotated towards the positive $z$-axis and becomes aligned along that direction as $f(Q_1) \rightarrow 0$ and if $f(Q_1)>1$, the spin is rotated away from the positive $z$-axis and becomes aligned opposite to that direction as $f(Q_1) \rightarrow \infty$. A NM corresponds to either $f(Q_1) = 0$ or $f(Q_1) = \infty$. The angle of rotation $\Delta \theta (\theta_{p}, Q_1)$ is shown in Fig.~1 as a function of $Q_1$ for a spin originally in the $xy$-plane ($\theta_{p}=\pi/2$) and the WMA prepared in a Gaussian state with width $\Delta_{\psi}=a$. 

\begin{figure}
\includegraphics[scale=0.55]{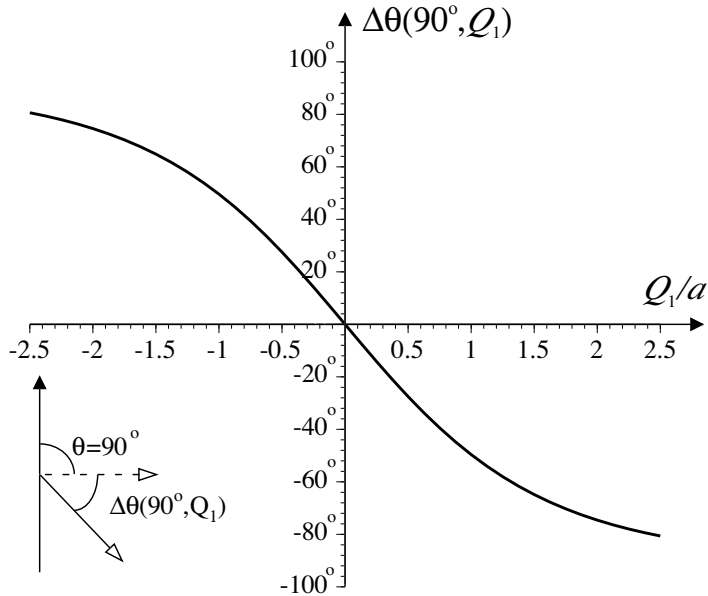}
\caption{A spin is prepared in a direction making angle $\theta_{p}$ with respect to the positive $z$-axis which is a direction set by a weak measurement apparatus. The pointer of the weak measurement apparatus originally has a Gaussian wave function of width $a$. After the weak measurement interaction the pointer is observed to have the position $Q_1$.  As a result, the direction of the spin is rotated through an angle $\Delta \theta(\theta_{p},Q_1)$ away from the $z$-axis. The dependance of the angle of rotation with $Q_1$ is shown for the case when the original direction $\theta_{p}=90^{\circ}$.}
\end{figure}

\section{Weak measurements and HV theories for entangled states}

\subsection{CNL HV model}
We will consider the CNL theory of Ref.~\onlinecite{leggett} for spins rather than the photons of the original formulation (the equivalence of the two cases was pointed out in Ref.~\onlinecite{leggett}). In the theory, instead of emitting pairs of spins in an entangled state $\Psi$, the source emits pairs with spins in directions $\hat{\bm{u}}$ and $\hat{\bm{v}}$ respectively with probability density $F(\hat{\bm{u}},\hat{\bm{v}})$ in the four-dimensional space $UV$ of the spin directions. The measurement outcomes on each pair of spins, including the nonlocal correlations, are controlled by a HV $\lambda$ with probability density $g_{\hat{\bm{u}}\hat{\bm{v}}}(\lambda)$ in the space $\Lambda$ of the HVs. The probability densities  $F(\hat{\bm{u}},\hat{\bm{v}})$ and  $g_{\hat{\bm{u}}\hat{\bm{v}}}(\lambda)$ must satisfy
\begin{equation}
\label{eq:probnorm}
1 \geq g_{\hat{\bm{u}}\hat{\bm{v}}}(\lambda) \geq 0 \text{  and  }  1 \geq F(\hat{\bm{u}},\hat{\bm{v}}) \geq 0
\end{equation}
and 
\begin{equation}
\int_{\Lambda} d\lambda g_{\hat{\bm{u}}\hat{\bm{v}}}(\lambda) =1 \text{  and  }  \int_{U} d^2\hat{\bm{u}} \int_{V} d^2\hat{\bm{v}} F(\hat{\bm{u}},\hat{\bm{v}}) =1 .
\end{equation}
The spins will be identified as the left-hand side (lhs) and  right-hand side (rhs) spins. After time evolutions $\hat{U}_l(t)$ on the lhs and $\hat{U}_r(t)$ on the rhs (with $\hat{U}_l(t)\hat{U}_r(t)=\hat{U}(t)$ below), NMs are performed on the lhs spin in the $\hat{\bm{a}}$-direction and on the rhs spin in the $\hat{\bm{b}}$-direction. The eigenstates of the measurements will be denoted respectively by $|a_{\alpha} \rangle $ and $|b_{\beta} \rangle $, $\alpha, \beta = \pm1$, with corresponding eigenvalues $\alpha$ and $\beta$ in units of $\hbar/2$. We consider the probability $\text{Prob}_{\text{hv}}[a_{\alpha},b_{\beta}|\hat{U}(t)]$ of outcomes $\alpha$ and $\beta$ on the lhs and rhs respectively (the dependence on the preparation state $\Psi$ will be repressed). According to the CNL HV theory, the probability $\text{Prob}_{\text{hv}}[a_{\alpha},b_{\beta}|\hat{\bm{u}},\hat{\bm{v}},\hat{U}(t)]$ of the outcomes $\alpha$ and $\beta$ given that the source emits spins in the directions $\hat{\bm{u}},\hat{\bm{v}}$ is given by
\begin{equation}
\label{eq;secondprob}
\text{Prob}_{\text{hv}}[a_{\alpha},b_{\beta} |\hat{\bm{u}},\hat{\bm{v}},\hat{U}(t)] = \int_{\Lambda}\text{Prob}_{\text{hv}}[a_{\alpha},b_{\beta} |\hat{\bm{u}},\hat{\bm{v}},\lambda(t),\hat{U}(t)] g_{\hat{\bm{u}}\hat{\bm{v}}}(\lambda(t),t)d\lambda(t).
\end{equation}
It is a requirement of the theory \cite{leggett} that for the marginal probabilities on the lhs and rhs (but not the joint probability), the $\hat{\bm{u}},\hat{\bm{v}}$ spins behave ``normally" when averaged over the HV $\lambda$. That is, for the lhs, the marginal probability must give the quantum mechanical result $\text{Prob}_{\text{qm}}[a_{\alpha}|\hat{\bm{u}},\hat{U}(t)]$ for outcome $\alpha$ of a measurement in the $\hat{\bm{a}}$ direction of a spin in the $\hat{\bm{u}}$ direction (which corresponds to the state $|u_+ \rangle$ in the present notation)
\begin{equation}
\label{eq;qmsecondcon}
\text{Prob}_{\text{hv}}[a_{\alpha} |\hat{\bm{u}},\hat{\bm{v}},\hat{U}(t)] =   \text{Prob}_{\text{qm}}[a_{\alpha}|\hat{\bm{u}},\hat{U}(t)] = |\langle a_{\alpha}|\hat{U}_l(t)|u_+ \rangle|^2.
\end{equation}
This will be called the secondary condition in the following. Of course, in order to agree with experiment, when averaged over $\hat{\bm{u}},\hat{\bm{v}}$, both the joint and marginal probabilities must give the quantum mechanical result. For the marginal probability on the lhs, this means
\begin{equation}
\label{eq;primecon}
\int_{UV}\text{Prob}_{\text{hv}}[a_{\alpha} |\hat{\bm{u}},\hat{\bm{v}},\hat{U}(t)] F(\hat{\bm{u}},\hat{\bm{v}}) d^2\hat{\bm{u}}d^2\hat{\bm{v}} = \text{Prob}_{\text{qm}}[a_{\alpha}|\hat{U}(t)]= |\langle a_{\alpha}|\hat{U}_l(t)|\Psi \rangle|^2 .
\end{equation}
This will be called the primary condition. Combining the primary and secondary conditions in Eqs.~(\ref{eq;qmsecondcon}) and (\ref{eq;primecon}) means that
\begin{equation}
\label{eq;primesecond}
\text{Prob}_{\text{hv}}[a_{\alpha}|\hat{U}(t)] = \int_{UV}|\langle a_{\alpha}|\hat{U}_l(t)|u_+ \rangle|^2 F(\hat{\bm{u}},\hat{\bm{v}}) d^2\hat{\bm{u}}d^2\hat{\bm{v}} = |\langle a_{\alpha}|\hat{U}_l(t)|\Psi \rangle|^2. 
\end{equation}

If $\hat{U}_l(t)$ is unitary, the transformation of $|u_+ \rangle$ in its two-dimensional spin space can be replaced by a rotation of the direction of measurement $\hat{\bm{a}}$ in three-dimensional space \cite{redhead}. Since the primary and secondary requirements must already be satisfied for any direction $\hat{\bm{a}}$, unitary time-evolution imposes no further condition on the HV theory. In the next section, we show that including the non-unitary time-evolution operator involving a WM does impose a further condition on the HV theory.

\subsection{WM and the HV model}
In order to investigate the CNL HV theory using WMs, it is sufficient to consider the experimental set-up shown in Fig.~2. Singlet states are produced by the source S and the lhs subsystem is subjected first to a rotation $R_x(\alpha)$ through an angle $\alpha$ about an axis specifying the $x$-axis and then by a WM whose orientation specifies the $z$-axis. The time-evolution $\hat{U}_l(t)$ operator for the WM is given by Eq.~(\ref{U}). It is also sufficient for the present purposes to consider the case when the final NM on the lhs is along the $z$-axis (the direction specified by the WMA). The rhs subsystem is measured in the direction $\hat{\bm{b}}$.

\begin{figure}
\includegraphics[scale=0.55]{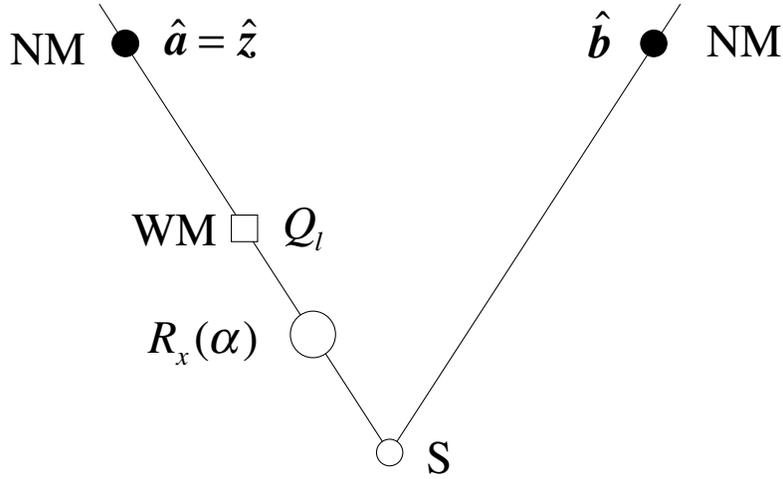}
\caption{A singlet state is produced by source S. The left-hand side subsystem is subjected first to a rotation through an angle $\alpha$ about the $x$-axis and then a weak measurement along the $z$-axis and finally a normal measurement also along the $z$-axis. For the right-hand side subsystem, a normal measurement is performed in the $\hat{\bm{b}}$-direction. The result of the normal measurement of the pointer position of the weak measurement apparatus is $Q_l$.}
\end{figure}

The quantum systems and the WMA are prepared at $t=0$ in the state
\begin{equation}
\label{eq;initstate}
| \Psi \rangle = \frac{1}{\sqrt{2}} (| z_+ \rangle_l  | z_- \rangle_r | - | z_- \rangle_l | | z_+ \rangle_r)|\psi_l\rangle  
\end{equation}
where $|\psi_l \rangle$ is the initial state of the WMA. We will be concerned with probabilities conditional on the outcome $Q_l$ of the WM (the WMs with different outcomes can be considered to be discarded).

If, according to the CNL HV model, the source emits a spin oriented in the direction $\hat{\bm{u}}$ specified by the polar angles $\theta_u,\phi_u$, after the rotation the spin is oriented in the direction specified by the polar angles $\theta_u(\alpha), \phi_u(\alpha)$ where
\begin{equation}
\label{eq;dangles}
\theta_u(\alpha) = \arccos(\cos \theta_u \cos \alpha + \sin \theta_u \sin \phi_u \sin \alpha)
\end{equation}
If the outcome of the WM pointer-position is $Q_l$, then from Eq.~(\ref{eq;tangamma}), the direction of the spin with respect to the $z$-axis becomes
\begin{equation}
\theta_l = 2 \arctan\left[ f(Q_l)\tan \frac{\theta_u(\alpha)}{2}\right].
\end{equation}
Therefore, for the specified time-evolution, the probability involved in the secondary requirement in Eq.~(\ref{eq;qmsecondcon}) is
\begin{equation}
\label{eq;secondfinal}
\text{Prob}_{\text{qm}}[a_{\alpha}|\hat{\bm{u}},\hat{U}(t)] = |\langle z_+|\hat{U}_l(t)|u_+ \rangle|^2 =  \cos^2\left( \arctan\left[ f(Q_l)\tan \frac{\theta_u(\alpha)}{2}\right] \right).
\end{equation}
A straightforward calculation using Eq.~(\ref{eq;initstate}) leads to the primary requirement for this case, i.e. the marginal probability (conditional on $Q_l$) for the outcomes spin-up in the $\hat{\bm{z}}$-direction on the lhs,
\begin{equation}
\label{eq;primefinal}
\text{Prob}_{\text{qm}}[a_{\alpha}|\hat{U}(t)]= |\langle z_+|\hat{U}_l(t)|\Psi \rangle|^2 = \frac{1}{1+(f(Q_l))^2}.
\end{equation}
From Eqs.~(\ref{eq;primesecond}), (\ref{eq;secondfinal}) and (\ref{eq;primefinal}), the primary and secondary requirements of the HV theory for this particular case mean that
\begin{equation}
\label{eq;finalcondition}
\int_{UV}\cos^2 \left[ 2 \arctan \left( f(Q_l)\tan \frac{\theta_u(\alpha)}{2} \right ) \right] F(\hat{\bm{u}},\hat{\bm{v}}) d^2\hat{\bm{u}}d^2\hat{\bm{v}} = \frac{1}{1+f(Q_l)^2}.
\end{equation}
From Eq.~(\ref{eq;dangles}), $\theta_u(\alpha)$ depends on $\alpha$ which can be chosen arbitrarily, which means that Eq.~(\ref{eq;finalcondition}) cannot be satisfied in general. Therefore the time-evolution involving the WM rules out the CNL HV.  The previous argument \cite{leggett,groblacher,branciard} that the CNL theory was not viable was that the CNL could not satisfy certain inequalities involving results for several combinations of $\hat{\bm{a}}$ and $\hat{\bm{b}}$. The present demonstration is simpler and more direct because it involves results for one value of $\hat{\bm{a}}$ and is independent of $\hat{\bm{b}}$. 

\section{Conclusion}

The main result is that a WM causes a change in a quantum system to a new state which depends on the prior state of the quantum system and the (uncontrollable) measured position of the pointer variable of the WMA. This provides a way of manipulating a state which is significantly different from a NM because, after a NM, the new state of the quantum system is one of the eigenstates of the observable that is measured is therefore independent of the prior state of the quantum system. In a NM the only role of the prior state is to influence the probabilities with which the eigenstates of the observable are taken up. The rotation of the spin subjected to a WM interaction has been discussed before \cite{botero} but in terms of the operator conjugate to the pointer-variable operator $\hat{Q}$ and not, as here, in terms of the WM outcome $Q$. It is significant that the WM, together with a NM of the pointer variable of the WMA, causes a non-unitary evolution of the quantum system. 

It follows that consideration of a WM of a quantum system can impose a significant constraint on possible HV formulations of quantum mechanics. This is because a HV theory may assign states to a quantum system which are different from the quantum mechanical state. A unitary transformation changes all states in a way which is independent of the prior state and so, if the HV theory mimics standard quantum mechanics before the unitary transformation, it will continue to do so afterwards.  In contrast, the WM will change the HV states in a way which depends on those assigned states and will also change the quantum mechanical state by an amount which depends on it.  The extra constraint is that the HV theory must continue to mimic standard quantum mechanics both with and without those disparate changes due to the WM. In Sec.~IIIB it was shown that those constraints rule out the setting-dependent CNL HV model proposed by Leggett \cite{leggett} in a more direct way than previously.

\end{document}